\documentclass{stile/aa}

\usepackage{txfonts}
\usepackage[english]{babel}
\usepackage{amsmath,amssymb}
\usepackage{graphicx, subfig}
\usepackage[normalem]{ulem}

\usepackage{verbatim}

\usepackage{multirow}
\usepackage{mathtools}
\usepackage{epstopdf}

\usepackage{url}
\usepackage{xcolor,color}
\usepackage{orcidlink}

\usepackage{natbib,twoopt}
\usepackage[hyphenbreaks]{breakurl}

\usepackage{tabularx}

\bibpunct{(}{)}{;}{a}{}{,}             %% natbib format for A&A and ApJ
\definecolor{cobalt}{rgb}{0.06, 0.2, 0.65}
\hypersetup{
  colorlinks,
  citecolor=cobalt,
  linkcolor=[rgb]{0.8, 0.2, 1.0},
  urlcolor=cobalt,
}
\makeatletter
  \newcommandtwoopt{\citeads}[3][][]{\href{http://adsabs.harvard.edu/abs/#3}%
    {\def\hyper@linkstart##1##2{}%
     \let\hyper@linkend\@empty\citealp[#1][#2]{#3}}}
  \newcommandtwoopt{\citepads}[3][][]{\href{http://adsabs.harvard.edu/abs/#3}%
    {\def\hyper@linkstart##1##2{}%
     \let\hyper@linkend\@empty\citep[#1][#2]{#3}}}
  \newcommandtwoopt{\citetads}[3][][]{\href{http://adsabs.harvard.edu/abs/#3}%
    {\def\hyper@linkstart##1##2{}%
     \let\hyper@linkend\@empty\citet[#1][#2]{#3}}}
  \newcommandtwoopt{\citeyearads}[3][][]%
    {\href{http://adsabs.harvard.edu/abs/#3}
    {\def\hyper@linkstart##1##2{}%
     \let\hyper@linkend\@empty\citeyear[#1][#2]{#3}}}
\makeatother

\def\be{\begin{equation}}
\def\ee{\end{equation}}

\def\cc{{\rm cm}^{-3}}

\def\gnot{{\rm G}_{0}}
\def\msun{{\rm M}_{\odot}}
\def\zsun{{\rm Z}_{\odot}}

\def\msunpc2{\msun/{\rm pc}^{2}}

\def\gsim{\lower.5ex\hbox{\gtsima}} 
\def\lsim{\lower.5ex\hbox{\ltsima}} 
\def\gtsima{$\; \buildrel > \over \sim \;$} 
\def\ltsima{$\; \buildrel < \over \sim \;$} \def\gsim{\lower.5ex\hbox{\gtsima}} 
\def\lsim{\lower.5ex\hbox{\ltsima}} 
\def\simgt{\lower.5ex\hbox{\gtsima}} 
\def\simlt{\lower.5ex\hbox{\ltsima}}
\def\nh2{n_{\rm H2}}

\definecolor{mkcolor}{HTML}{01abdf} %mahsa
\definecolor{apcolor}{HTML}{b3003b}
\definecolor{afcolor}{HTML}{01bdff}
\definecolor{blue-violet}{rgb}{0.54, 0.17, 0.89}
\definecolor{ao}{rgb}{0.0, 0.5, 0.0}
\definecolor{auburn}{rgb}{0.43, 0.21, 0.1}

\graphicspath{{plots/},{plots_appendix}}

\begin{document}

\title{Emulating the interstellar medium chemistry with neural operators}
\titlerunning{Emulating ISM chemistry with DeepONet}

\author{
Lorenzo Branca \orcidlink{0000-0002-6064-1964} \inst{1}\fnmsep\thanks{\href{mailto:lorenzo.branca@sns.it}{lorenzo.branca@sns.it}} \and
Andrea Pallottini \orcidlink{0000-0002-7129-5761} \inst{1}.
     }
\authorrunning{Branca \& Pallottini}
\institute{Scuola Normale Superiore, Piazza dei Cavalieri 7, 56126 Pisa, Italy}
\date{Received 9 January, 2024; accepted XX XX, XXXX}

\abstract
% context
{Galaxy formation and evolution critically depend on understanding the complex photo-chemical processes that govern the evolution and thermodynamics of the InterStellar Medium (ISM). Computationally, solving chemistry is among the most heavy tasks in cosmological and astrophysical simulations.
}
% aims
{Astrophysical simulations can include photo-chemical models that allow for a wide range of densities ($n$), abundances of different species ($n_i/n$) and temperature ($T$), and evolving the ISM under the action of a radiation field ($\textbf{F}$) with different spectral shape and intensity. The evolution of such non-equilibrium photo-chemical network relies on implicit, precise, computationally costly, ordinary differential equations (ODE) solvers. Here, we aim at substituting such procedural solvers with fast, pre-trained, emulators based on neural operators.
}
% method
{We emulate a non-equilibrium chemical network up to H$_2$ formation (9 species, 52 reactions) by adopting the DeepONet formalism, i.e. by splitting the ODE solver operator that maps the initial conditions and time evolution into a tensor product of two neural networks (named branch and trunk).
We use $\texttt{KROME}$ to generate a training set spanning $-2\leq \log(n/\cc) \leq 3.5$, $\log(20) \leq\log(T/\mathrm{K}) \leq 5.5$, $-6 \leq \log(n_i/n) < 0$, and by adopting an incident radiation field $\textbf{F}$ sampled in 10 energy bins with a continuity prior.
We separately train the solver for $T$ and each $n_i$ for $\simeq 4.34\,\rm GPUhrs$.
}
% results
{Compared with the reference solutions obtained by $\texttt{KROME}$ for single zone models, the typical precision obtained is of order $10^{-2}$, i.e. the $10 \times$ better with a training that is $40 \times$ less costly with respect to previous emulators which however considered only a fixed $\mathbf{F}$.
DeepONet well behaves also for $T$ and $n_i$ outside the range of the training sample.
Further, the emulator well reproduces also the ion and temperature profiles of photo dissociation regions, i.e. giving errors comparable to the typical difference between various photo-ionization codes.
The present model achieves a speed-up of a factor of $128 \times$ with respect to stiff ODE solvers.
}
% conclusions
{Our neural emulator represents a significant leap forward in the modeling of ISM chemistry, offering a good balance of precision, versatility, and computational efficiency.
Nevertheless, further work is required to address the challenges represented by the extrapolation beyond the training time domain and the removal of potential outliers.
}


\maketitle

\section{Introduction}

%
% intro sentence
%
Non-equilibrium photo-chemistry plays a crucial role in many astrophysical and cosmological environments.
%
% usage cases for photo-chemistry
%
Chemistry regulates the physical processes starting from cosmological scales \citep{galli:1998,glover:2008}, is key to understand the evolution of the Intergalactic Medium (IGM) during the Epoch of Reionization \citep[EoR,][]{theuns:1998, maio:2007, rosdahl:2018}, has a large impact in the formation and evolution of galaxies \citep{pallottini:2017_b, lupi:2019}, in particular regulating the birth of Giant Molecular Clouds \citep[GMC,][]{decataldo:2019, kim:2018} in the InterStellar Medium (ISM), and continue to play a vital role down to the formation of planets \citep{caselli:2012}.

%
% how to addres the problem
%
From a theoretical point of view, these widely different environments are studied by developing astrophysical and cosmological simulations. In order to follow the thermo-chemical evolution in such simulations, it is necessary to solve the system of Ordinary Differential Equations (ODEs) associated with it. 
Depending on the specific problem, various software have been developed and used to include chemical non-equilibrium chemistry in numerical codes.
%
% accurate/slow
%
On the one hand, some codes allow for extensive photo-chemical network, are very precise, thus very consequently expensive, such as \texttt{CLOUDY} \citep{ferland:2017}, \texttt{UCLCHEM} \citep{holdship:2017}, and \texttt{MAIHEM} \citep{gray:2019}; this kind of approach \citep[see][for a review]{olsen:2018} is typically more suitable for post-processing astrophysical simulations to obtain emission lines \citep[e.g.][]{vallini:2018}, since on-the-fly direct coupling is commonly limited to 1D simulations, such as photo-chemical evolutionary patterns that are solved during shock processes \citep[e.g.][]{danehkar:2022}.
%
% hydro coupling
%
On the other hand, some software are designed to be coupled to code for full 3D hydrodynamic evolution, such as \texttt{ASTROCHEM} \citep{kumar:2013MNRAS} and \texttt{KROME} \citep{grassi:2014}, which are interfaces that, by taking as input an user defined chemical network, are used to prepare the code needed to solve the associated ODE system, or as \texttt{NIRVANA} \citep{ziegler:2016} and \texttt{GRACKLE} \citep{smith:2017}, which are libraries providing subroutines to solve selected types of chemical networks that are of astrophysical interest.

%
% problems in coupling chemical and hydro evolution
%
Coupling the chemistry evolution with an astrophysical code is challenging, mainly since the chemical ODE system is often stiff and the typical time-scales are much shorter than the dynamical/hydrodynamic time, e.g. $\Delta t_{chem}\ll 10^{-4}\Delta t_{hydro}$: this implies that robust, multistep, embedded, high-order, and thus ultimately costly, numerical solvers should be selected \citep[e.g.][]{odepack}.
Further, in terms of number of reactions, the complexity of the system grows more than linearly with respect to the number of species included in the chemical network.
Finally, consider that in numerical simulations the domain is usually split on the basis of the time-steps hierarchy, in order to balance between different processors the cost of gravity evaluation and hydrodynamical computation \citep{springel:2021}, while it is difficult to estimate the time-to-completion of the chemistry solver given the initial conditions \citep{branca:2023}, implying the inclusion of such a task can spoil the load balancing.

%
% ML to overcome limitations
%
In order to overcome these limitations, various methods based on deep learning have been developed in recent years. 
%
% post-processing ML
%
In \citet{holdship:2021}, the \texttt{CHEMULATOR} algorithm is introduced in order to emulate \texttt{UCLCHEM} \citep{holdship:2017}; it utilizes a combination of autoencoders and Principal Component Analysis to reduce the dimensionality of the chemical network (from 33 down to 8 dimension); subsequently, a standard Feed-Forward Neural Network (FNN) is employed to predict the evolution of latent space variables.
For the same kind of complex chemical system, in \citep{heyl:2023} the interpretability of the network is showcased by applying the SHAP coefficient calculation \citep{lundberg:2018} to \texttt{UCLCHEM} pre-computed tables interpolated via the eXtreme Gradient Boosting \texttt{XGboost} \citep{XGboost:2016} regressor.
Further, machine learning techniques such as random forest can be used in conjunction with libraries of precise photoionization models in order to quickly predict the line emission from numerical simulations \citep{katz:2019} or infer the basic ISM properties from observed line ratios \citep{ucci:2018}.

%
% coupling ML
%
Going in the direction of a coupling with numerical simulation, \citet{robinson:2023} presents the feasibility of constructing an emulator utilizing \texttt{XGboost} for computing gas heating and cooling tables obtained from \citet{gnedin:2012}; this approach relies on accurate \texttt{CLOUDY} \citep{ferland:2017} models for the training, however does not account for the evolution of chemical species \citep[see also][for an emulator for cooling]{galligan:2019}.
Meanwhile, \citet{grassi:2021} investigates the application of autoencoders to simplify the chemical network's complexity and solve the associated system of reduced ODEs within the encoded latent space. This concept shows promise, especially for intricate and large chemical networks, such as those with over 400 reactions \citep{glover:2010}, suitable for molecular cloud/clump studies; however, the current implementation lacks support for temperature evolution and rate-dependent coefficients.
In \citet{branca:2023}, the possibility to use Physics Informed Neural Networks is explored; the method consists in preparing a neural network by embedding the residual of the ODE associated chemical system directly in the loss function (the physical informed part); while such a method can be regarded as elegant, since it does not require a pre-calculated training set, the training is costly ($\sim 2 \mathrm{kGPUhr}$ to reach a $10\%$ relative accuracy), thus it is difficult to achieve an accuracy level high enough to be competitive with procedural solvers.
%

%
% radiation dependence
%
Furthermore, an additional complication is given by the coupling between chemistry and radiation. Up to now, chemical emulators prepared for hydrodynamical coupling \citep{grassi:2021,branca:2023} allowed only for a fixed shape and intensity of the incident radiation. Instead, a more refined model should be able to be fully paired with an arbitrary radiation field, for instance expressed as a function of frequency, that is required for a coupling appropriate for radiation hydrodynamic simulations \citep[e.g.][]{rosdahl:2018, pallottini:2019, decataldo:2020, trebitsch:2021,katz:2022,2019MNRAS.490.1518O,2024MNRAS.527.8078O}.
However, this adds a further degree of complexity to the problem that is not fully explored; indeed, to our knowledge, the only example of such exploration is given by \citet{robinson:2023}, where cooling/heating tables from \citet{gnedin:2012} are emulated, and such tables are computed by assuming photo-ionization equilibrium conditions as a function of the incident radiation field; however, in \citet{gnedin:2012} the radiation is approximated by adopting a limited number of photo-ionization rates, and, by construction, the model cannot yield the evolution of the ISM chemistry.

% aim of the current work
To this end, a novel possibility consists in exploiting Neural Operators to emulate the photo-chemical ODEs system. The use of neural operators could be crucial in learning the differential operator that describes the chemical network, giving the emulator the flexibility required to be effectively used in simulations.
In particular, in this work we want to explore the usage of a particular architecture, Deep Neural Operator (DeepONet), firstly introduced in \citet{lu2021learning}. DeepONet is a versatile and robust tool that already showed good performance both for relatively simple case of studies, such as Diffusion-Reaction system \citep{lu2021learning}, and for more complex problems, such as hyper-sonic shocks \citep{2021JCoPh.44710698M}.

By exploiting the DeepONet architecture, here we aim at developing a model capable of improving the quality and efficiency of the results obtained in \citet{branca:2023}, i.e. \textit{i)} emulating a relatively complex chemical network (up to H$_2$ formation), \textit{ii)} allowing for a dependence from a general incident radiation, and \textit{iii)} improving upon the cost of reaching high precision with the training, in summary to have an efficient emulator ready to be coupled with hydrodynamical simulations that include radiative transfer on the fly.

\section{Method}

The aim of this work is to prepare an emulator for ISM photochemistry.
First we summarize the chemical network selected (Sec. \ref{sec:ISM_chemistry}), then we detail the implementation used for this work (Sec. \ref{sec:deeponet}), and finally we give the setup of the data-set used for the training of the emulator (Sec. \ref{sec:dataset}).

\begin{figure*}
    \centering
    \includegraphics[width=0.9\textwidth]{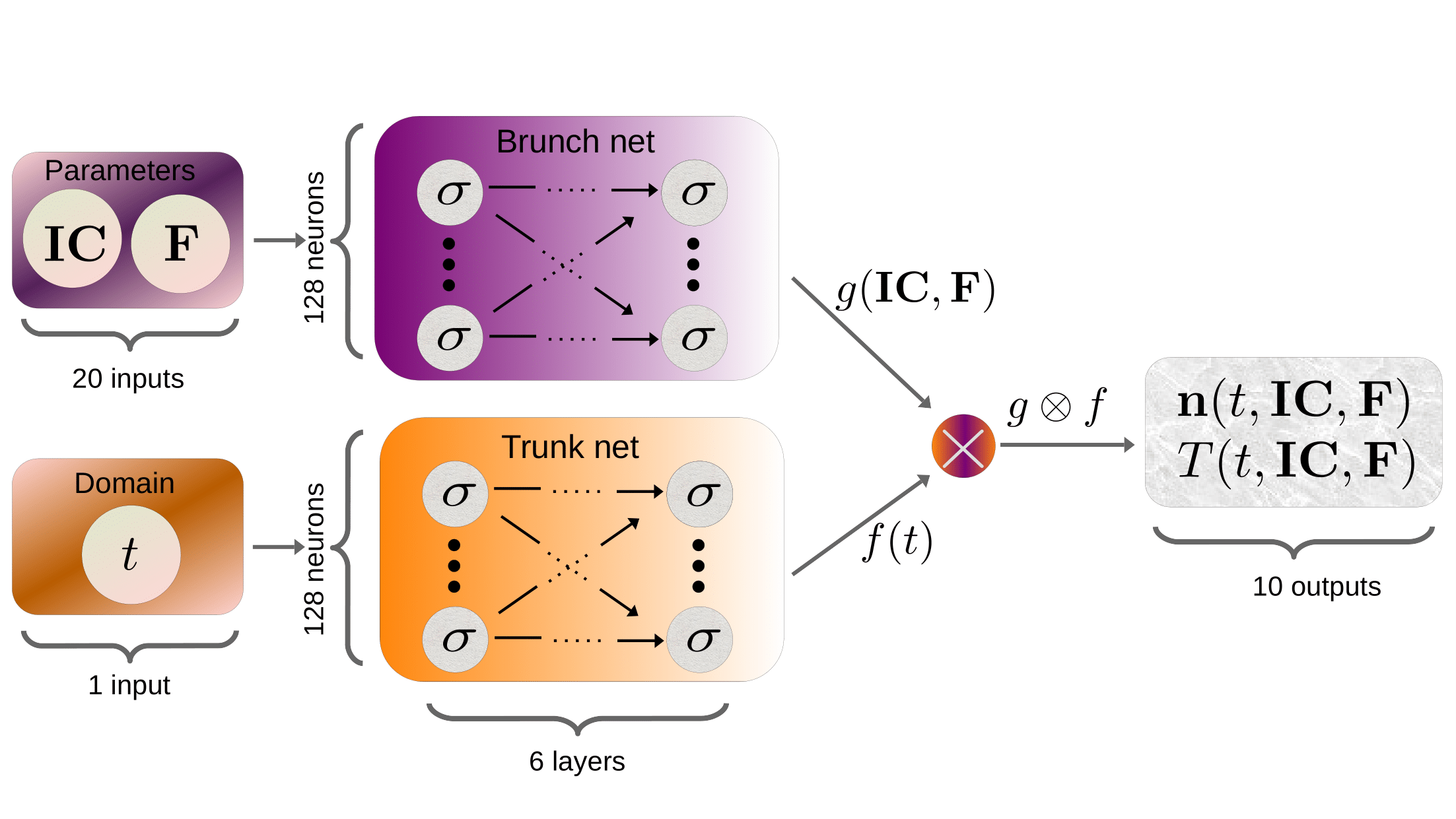}
    \caption{Scheme of the emulator implemented in this work.
    The system Ordinary Differential Equations (ODE) describing the InterStellar Medium (ISM) chemical network (Sec. \ref{sec:ISM_chemistry}) is emulated via the DeepONet formalism (Sec. \ref{sec:deeponet}) by splitting the dependence i) from the initial conditions ($T$, and densities $n$ of each chemical species, i.e. $\mathrm{e}^-$, $\mathrm{H}^-$, $\rm H$, $\mathrm{H}^+$, He, $\mathrm{He}^+$, $\mathrm{He}^{++}$, $\mathrm{H}_2$, and $\mathrm{H}_2^+$) and the radiation flux ($\mathbf{F}$) with the branch network ($g$) and ii) from the temporal evolution in the time ($t$) domain with a trunk network ($f$); $f$ and $g$ are feed-forward neural networks, each consisting of 6 dense layers with 128 neurons: the tensor product ($\otimes$) of the branch and the trunk is adopted to compute the loss function (eq. \ref{eq:UATfO}).
    We individually train networks for the temperature and each of the chemical species; the data-set adopted to train DeepONet is described in Sec. \ref{sec:dataset} and its main properties are summarized in Tab. \ref{tab:data_set_structure}.
    \label{fig:sketch_network}
    }
\end{figure*}

\subsection{ISM photo-chemistry}\label{sec:ISM_chemistry}

\begin{table}
\centering
\begin{tabular}{lllc}
\hline
\multicolumn{3}{l}{Photo-reaction} & $h \nu_{\rm min}/{\rm eV}$             \\
\hline
${\rm H}^- +\gamma $  & $\rightarrow$ & ${\rm H} + e$                     & 0.76 \\
${\rm H}_2^+ +\gamma$ & $\rightarrow$ & ${\rm H}^+ + {\rm H}$             & 2.65 \\
${\rm H}_2 +\gamma $ & $\rightarrow $ & ${\rm H} + {\rm H}\,\,$ (Solomon) & 11.2 \\
${\rm H}+\gamma $ & $\rightarrow $ & ${\rm H}^+ + e$                      & 13.6 \\
${\rm H}_2 +\gamma $ & $\rightarrow$ & $ {\rm H} + {\rm H}\,\,$ (direct)  & 14.2 \\
${\rm H}_2 +\gamma $ & $\rightarrow$ & $ {\rm H}_2^+ + e$                 & 15.4 \\ 
${\rm He}+\gamma $ & $\rightarrow $ & ${\rm He}^+ + e$                    & 24.6 \\
${\rm H}_2^+ +\gamma$ & $ \rightarrow$ & $ {\rm H}^+ + {\rm H}^+ + e$     & 30.0 \\
${\rm He}^+ +\gamma$ & $ \rightarrow$ & $ {\rm He}^{++} + e$              & 54.4 \\
\hline
\end{tabular}
\caption{List of photo-chemical reactions included in the chemical network adopted in the present work and corresponding minimum frequency ($\nu_{\rm min}$) of the photon for its activation. See Sec. \ref{sec:ISM_chemistry} for details.
\label{tab:photo_reactions}
}
\end{table}

%
% chemical network selection
%
Similarly to \citet{branca:2023}, we adopt the ISM photo-chemical network described in \citet{bovino:2016} and summarized below. As such a chemical network has been used both in studies on molecular cloud \citep[][]{decataldo:2019,decataldo:2020} and galaxy \citep[][]{pallottini:2017_b,pallottini:2019,pallottini:2022} scales, we deem it as a good example for kick-starting the coupling with astrophysical simulations.

%
% species and reactions
%
The network tracks the evolution of $\mathrm{e}^-$, $\mathrm{H}^-$, $\rm H$, $\mathrm{H}^+$, He, $\mathrm{He}^+$, $\mathrm{He}^{++}$, $\mathrm{H}_2$, and $\mathrm{H}_2^+$, which evolve according to 46 reactions\footnote{The reaction rates are taken from \citet{bovino:2016}: reactions 1 to 31, 53, 54, and from 58 to 61 in their tables B.1 and B.2, photo-reactions P1 to P9 in their Tab. 2.}, involving dust processes, i.e. H$_2$ formation on dust grains \citep{jura:1975}, photo-chemistry (see Tab. \ref{tab:photo_reactions}), and cosmic rays ionization. Similarly to \citet{branca:2023}, we consider only solar metallicity, abundances ($Z=Z_{\odot}$, \citealt{asplund:2009}), and dust to gas ratio ($f_d=0.3$, \citealt{hirashita:2002}).

%
% mathematical formulation
%
The chemical network includes 2-body reactions and interaction with an input radiation field $\mathbf{F}$ that quantifies the photon and cosmic ray flux in various energy bins, thus the evolution of the species can be written as
\begin{equation}
    \dot{n}_k= \sum_{i,j} A^{ij}_k n_i n_j+ \sum_{i} B_k^i n_i\,,
    \label{2body}
\end{equation}
where $A^{ij}_k=A^{ij}_k(T,\mathbf{n})$ are the temperature ($T$) dependent 2-body reaction coupling coefficients, $B^{i}_k=B^{i}_k(\mathbf{F})$ describe the photo-reactions rates, and the indexes range on all the 9 included species.
%
% radiation field
%
For the radiation field, we select a constant cosmic ray flux of $\xi=3 \times 10^{-17}\rm s^{-1}$, which is appropriate for the Milky-Way \citep{webber:1998}, and set the coupling with the gas adopting the \texttt{kida} database \citep{wakelam:2012}; for the photons, we select 10 radiation bins such that all the 9 photo-reaction included in the chemical network are fully covered (see Tab. \ref{tab:photo_reactions}); the spectral shape of the incident radiation is further detailed in Sec. \ref{sec:dataset}.

%
% temperature evolution
%
The system can be considered complete once $T$ is simultaneously evolved
\begin{equation}
    \label{temp_evolution}
    \dot{T}=\frac{(\gamma-1)}{k_b\sum_{i}n_i}(\Gamma-\Lambda)\,,
\end{equation}
where $k_b$ is the Boltzmann constant, $\gamma$ is the gas adiabatic index, $\Gamma=\Gamma(T,\mathbf{n},\mathbf{F})$ and $\Lambda=\Lambda(T,\mathbf{n},\mathbf{F})$ are the heating and cooling functions, respectively.
%
% heating and cooling processes
%
$\Gamma$ includes contribution from photoelectric heating from dust \citep{bakes:1994}, cosmic rays \citep{cen:1992}, and photo-reactions (Tab. \ref{tab:photo_reactions}).
$\Lambda$ accounts for cooling from atoms \citep{cen:1992}, molecules (only molecular hydrogen here, \citet{glover:2008}), metal lines \citep{shen:2013}, and the Compton effect.
Additionally, exothermic and endothermic chemical reactions give contributions to the heating and cooling terms, respectively.

\subsection{Deep Operator Network for photo-chemistry}\label{sec:deeponet}

%
% UAT
%
Thanks to the Universal Approximation Theorem (UAT, \citealt{citeulike:1989}), a neural network can approximate any continuous function. Actually, a more general result has been demonstrated by \citet{chen2:1995}, the so called UAT for Operators.
%
% UATO
%
In principle, such extension allows neural networks to be used not only as function approximator, but also to learn maps between functional spaces, in particular, allowing for a family of functions to be approximated.
%

%
% practical applications of UATO
%
However, only years after the proof of UAT for operator has been given, the extended theorem yielded practical applications, as in \citet{lu2021learning}, where the authors presented DeepONet, an architecture capable of exploiting the idea of the original theoretical result from \citet{chen2:1995} in order to solve 2D Riesz fractional Laplacian, nonlinear diffusion-reaction PDEs with a source term, and Stochastic PDEs for population growth model.
Note that DeepONet is not the only proposed implementation for neural operators. Another very popular application is the Fourier Neural Operator (FNO), originally proposed in \citet{li:2020}; a comparison between the two architecture can be found in \citet{lu:2022}, where the authors also show the ability for neural operators to solve a one-dimensional Euler equation coupled with a simple three-species chemical network.
%

% 
% DeepONet
%
For this work, we adopt the DeepONet implementation, which is fully described in \citet{lu2021learning} and can be summarized as follows \citep[see also][]{lu:2019}.
Our task consists in emulating the operator $\mathcal{G}$, which maps the variable in the domain $\mathbf{x}$ to $\mathcal{G}(\mathbf{p})(\mathbf{x})$, depending on the vector $\mathbf{p}$ which contains the so-called \textit{sensor}, which, in general, can be functions.
%
% DeepONet for photo-chem
%
In our case, $\mathcal{G}$ is the ODE system describing the evolution of a photo-chemical network, and the domain $\mathbf{x}$ is limited to time; a family of solutions for our ODEs system is well defined once the initial conditions are given, thus the sensor $\mathbf{p}$ includes i) the initial temperature and density of the chemical species ($\mathbf{IC}$) and ii) the photon flux as a function of frequency ($\mathbf{F}$).
As illustrated in Figure \ref{fig:sketch_network}, the idea of DeepONet consist in splitting in two the emulation, feature that can be formalized by defining the loss function as
\begin{equation}
    \label{eq:UATfO}
    \mathcal{L} = \left\lVert
    \mathcal{G}(\mathbf{p})(\mathbf{x}) -  g(\mathbf{p})\otimes f(\mathbf{x})
    \right\rVert < \epsilon\,,
\end{equation} 
where $\Vert$ denotes a yet to be defined norm, $\epsilon$ is a target value for the training, $\otimes$ denotes the tensor product,  the functions $g$ and $f$ represent the \textit{branch} and \textit{trunk} neural networks, respectively and $\mathcal{G}(\mathbf{g})(\mathbf{x})$ is the pre-computed ground truth solution (see later Sec. \ref{sec:dataset}). The inequality in eq. \ref{eq:UATfO} is well motivated precisely by the universal approximation theorem for operators.

We developed our emulator using \texttt{DeepXDE} \citep{DeepXDE:2019} with the \texttt{TensorFlow} backend \citep{tensorflow:2015}, a \texttt{python} package \citep{python3} that provides a high-level application programming interface for the implementation of deep learning methods both for solving forward and inverse problems that can be described by differential equations; other tools designed for this purpose are \texttt{MODULUS} \citep{hennigh2020nvidia} that is provided by \texttt{NVIDIA} and \texttt{sciML} \citep{sciml_diffeqflux_2019} which is written for \texttt{julia} \citep{julia}. 

The two feed-forward neural networks, referred to as branch and trunk, each consist of 6 dense layers with 128 neurons, employing the Rectified Linear Unit (ReLU) activation function. We initialize the weights using the Glorot normal prescription \citep{Glorot:2010} , that is considered a standard\footnote{See \url{https://github.com/lululxvi/deeponet/tree/master}.} for this kind of architectures \citep[][]{lu2021learning}, set the initial learning rate to $lr=10^{-3}$, and utilize the ADAM optimization algorithm with default hyper-parameters initialization \citep{kingma:2014}.
To mitigate overfitting on the training data, we have incorporated a regularization technique; specifically, we augment the loss function (eq. \ref{eq:UATfO}) with two additional terms, implementing L1 and L2 regularization techniques.
We transform all the quantities in a non-dimensional logarithmic space such as:
\begin{equation}\label{eq:def_sum_ion_and_t}
y =  2\frac{x_i - x^{min}_i}{x^{max}_i - x^{min}_i} - 1\, ,
\end{equation}
where $i$ ranges on all ion densities and $T$, $x_i = \log (n_i / \rm cm^{-3})$ and $x_i = \log (T / \rm K)$, respectively.

It is important to point out that, to approximate the differential operator $\mathcal{G}$, we pre-compute a sub-set of solutions to use as training data, which makes this method data-driven (see Sec. \ref{sec:dataset}). Furthermore, a specific emulator is dedicated to each different chemical species (and temperature). This choice allows us to have better sampling in the space of the initial parameters, as here the training on a single species requires ten times less GPU memory than the multi-output model.

\subsection{Data-set generation}\label{sec:dataset}

\begin{table}
\centering
\begin{tabular}{l|llll }
 \hline
 quantity         & variable         & bins & min & max \\ 
 \hline
 gas density      & $\log (n/\cc)$   & 64   & -2 & 3.5 \\  
 abundances       & $\log (n_i/n)$   & 512  & -6 & 0   \\  
 temperature      & $\log (T/\rm K)$ & 64   & $\log(20)$  & 5.5\\    
 radiation        & $\log(F_i/\rm eV\,cm^{-2}\, s^{-1}\, Hz)$   & 64   & -15 & -5 \\
 time             & $t/\rm kyr$      & 16   & 0 & 1 \\
 \hline
\end{tabular}
\caption{Summary of the properties used to generate the training set for this work. 
As detailed in Sec. \ref{sec:dataset}, we vary the initial gas density ($n$), species abundances ($n_i/n$), and temperature ($T$), by allowing a for a radiation field split in 10 energy bins ($F_i$) and by solving the chemical evolution up to a time $t$.
Each model is generated with \texttt{KROME} for a chemical network up to H$_2$ formation (9 species, reactions see Sec. \ref{sec:ISM_chemistry} for details.)
In total, the data-set contains $644245094 \simeq 6.4 \times 10^{8}$ models.
\label{tab:data_set_structure}
}
\end{table}

%
% krome for photo-chem
%
The data used as train and test sets for our models are produced using \texttt{KROME} following the model described in Sec. \ref{sec:ISM_chemistry}.
\texttt{KROME}\footnote{\url{https://bitbucket.org/tgrassi/krome/src/master/}} \citep{grassi:2014} is a \texttt{python} interface that generates the \texttt{fortran} code to solve an input chemical model.
The associated ODE system is evolved in time by adopting a 5$^{th}$ order version of \texttt{LSODES} \citep{odepack}, an implicit, robust, multistep, iterative\footnote{We adopt the default errors for \texttt{KROME}, i.e. relative and absolute tolerances are fixed at $10^{-4}$ and $10^{-20}$ respectively.} solver that takes advantage of the sparsity of the Jacobian matrix contructed with the chemical fluxes.
%
% IC for n and T
%
For the gas number density $n$ and initial temperature $T$ we  randomly sample\footnote{In principle, the validity of the UAT for operators does not depend on the distribution of the training data. For DeepONet, a uniform random sampling is generally adopted in the literature; see \citet[][in particular see theorem 3.1]{2022arXiv220511404P} for an example of non-uniform data sampling.} 64 points in the log-spaced range $-2\leq \log(n/\cc) \leq 3.5$ and $\log(20) \leq\log(T/\mathrm{K}) \leq 5.5$, respectively, by extracting the points within the variables range.
Similarly, the initial fractions of each species $n_i/n$ are log-sampled in 512 bins in the range\footnote{Note that the limit on the fraction is applied only to the initial conditions, because of the chemical evolution individual species are allowed to be as low down to $n_{\rm min}=10^{-10}\cc$. } $-6 \leq \log(n_i/n) < 0$; we add the constraints that the total Hydrogen and Helium abundances are close to primordial and that global charge neutrality is respected.

%
% radiation bins
%
The radiation field $\mathbf{F}$ is composed by 10 energy bins for the photons (see Tab. \ref{tab:photo_reactions}) and is treated as follows\footnote{Note that in \texttt{KROME}, the radiation field is a pure input, i.e. individual fluxes $F_i$ are not directly evolved by the system. In the typical usage case, \texttt{KROME} is coupled to a numerical simulation, the former provides the opacities, the latter include the radiative transfer modules for the evolution of $F_i$ \citep[e.g.][]{pallottini:2019}.}.
For a full generalization, each component of the incident flux $F_i$ should be treated similarly to the other inputs ($n$, $n_i$, and $T$), i.e. randomly extracting the value in a pre-established range. However, this approach would likely generate a completely non-physical spectrum in the vast majority of cases: difference between contiguous $F_i$ would be unbound, e.g. a fully random spectrum might contain He ionizing photons and have a negligible flux in the \citet{habing:1968} band ($6.0-13.6\,\rm eV$).
Further, the sampling of $F_i$ should be as refined as the one for the relative abundances, which would increase the size of the data-set, making it harder to load it efficiently in the GPU memory.

We therefore adopted the following method: first we generated 64 flux values $F_{i}$ in any of the bins in a range $F_{i} \in [10^{-15}, 10^{-5}] \mathrm{eV/cm^2/s/Hz}$. Then we extracted a flux value in the adjacent bin such that $|\log(F_{i}/F_{i\pm 1})|\leq 0.15$. In this way the stochasticity of the incoming flows is maintained, without having completely non-physical spectra.
In this way we greatly reduced the size of the dataset, thus being able to refine the sampling of the initial conditions on the fractions.
Note that imposing such a constraint for the flux is somewhat similar to the selection of a continuity prior in spectral energy density fitting, when non-parametric star formation rate histories are explored by allowing for burstiness \citep{tacchella:2020,ciesla:2023}.
%

%
% time evolution
%
The time evolution is evaluated in 16 random uniformly sampled points in a range $0 \leq t/\mathrm{kyr} \leq 1$. In total, the training set comprises 30\% of the dataset and the remaining 70\% is used for on-the-fly (during the training) validation, totaling $644245094 \simeq 6.4 \times 10^{8}$ points. A summary of the adopted parameters is provided in Table \ref{tab:data_set_structure}.

\section{Results}

\begin{table}
\centering
\begin{tabular}{ l | c c c c  }
\hline
 input    & MRE & 50\% & 75\% & 90\% \\ 
 \hline
 $\mathrm{T}$       & 0.0179 & 0.0114 & 0.0211 & 0.0353\\  
 $\mathrm{e}^-$     & 0.0074 & 0.0044 & 0.0078 & 0.0151 \\  
 $\mathrm{H}^-$     & 0.0176 & 0.0076 & 0.0183 & 0.0398\\
 $\mathrm{H}$       & 0.0269 & 0.0190 & 0.0375 & 0.0547 \\
 $\mathrm{He}$      & 0.0213 & 0.0150 & 0.0258 & 0.0422 \\
 $\mathrm{H}_2$     & 0.0274 & 0.0167 & 0.0292 & 0.0548 \\
 $\mathrm{H}^+$     & 0.0099 & 0.0060 & 0.0101 & 0.0178 \\
 $\mathrm{He}^+$    & 0.0148 & 0.0082 & 0.0177 & 0.0314 \\
 $\mathrm{H}^+_2$   & 0.0255 & 0.0146 & 0.0267 & 0.0552 \\
 $\mathrm{He}^{++}$ & 0.0234 & 0.0125 & 0.0220 & 0.0436\\
 \hline
\end{tabular}
\caption{Quantitative summary of the distribution of relative errors, i.e. for the temperature and each chemical species emulated we report the Mean Relative Error (MRE) and the quantiles of the relative errors for 50\% (median), 75\%, and 90\% of the testing cases.
\label{tab:model_results}
}
\end{table}

\begin{figure}
    \centering
    \includegraphics[width=0.49\textwidth]{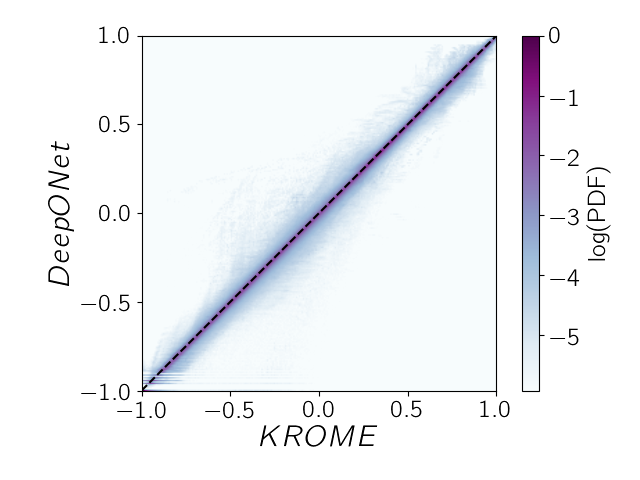}
    \caption{Predicted vs true test for the DeepONet model.
    Logarithmic densities for each ion (H, H$^{+}$, ...) and the temperature (T) are normalized in the full data-set range (Tab. \ref{tab:data_set_structure}) and summed ($y$, see eq. \ref{eq:def_sum_ion_and_t}) for both the true value from \texttt{KROME} and the predicted value from DeepONet.
    The image show the 2D probability distribution function (PDF) of the summed dataset, and it is normalized such that the maximum is 1 to better appreciate the dynamical range. To guide the eye, we have added a dashed black line to mark the $\rm KROME= DeepONet$ region.
    See Fig. \ref{fig:pred_vs_true_individual} for the same diagnostic for individual ions.
    \label{fig:pred_vs_true}
    }
\end{figure}

\begin{figure*}
    \centering
    \includegraphics[width=0.9\textwidth]{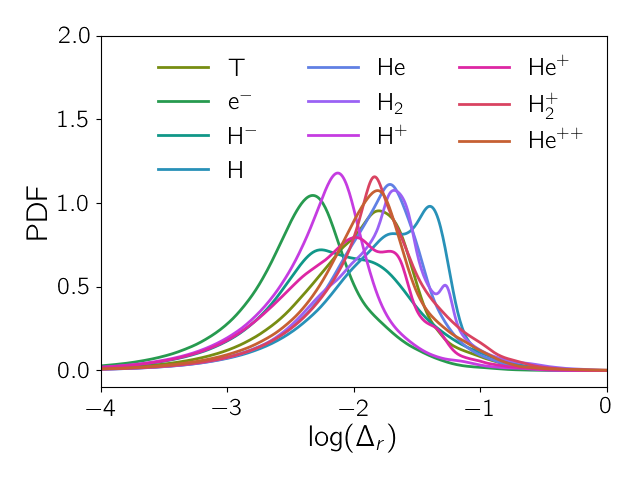}
    \caption{
    PDF of the relative error ($\Delta_r$) for the testing set.
    Each output from the emulator ($T$ and density for each species) is shown independently, with the color-code indicated in the legend.
    The PDFs are computed using a testing set of $\simeq 2 \times 10^7$ of points.
    % 20132656
    %
    \label{fig:error_pdf_1D}
    }
\end{figure*}

\begin{figure*}
    \centering
    \includegraphics[width=0.49\textwidth]{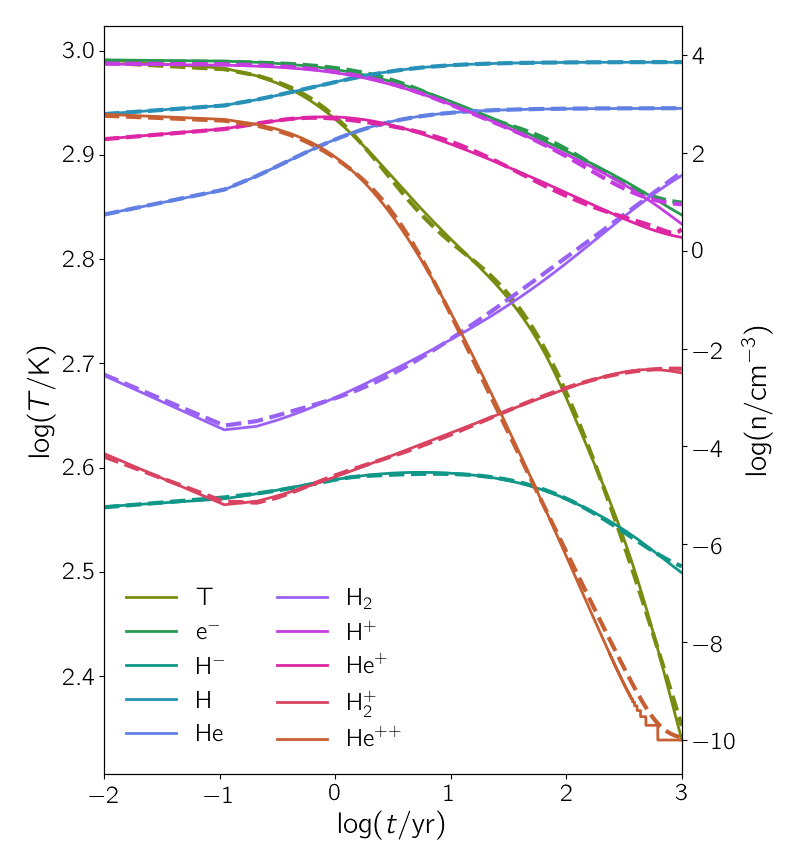}
    \includegraphics[width=0.49\textwidth]{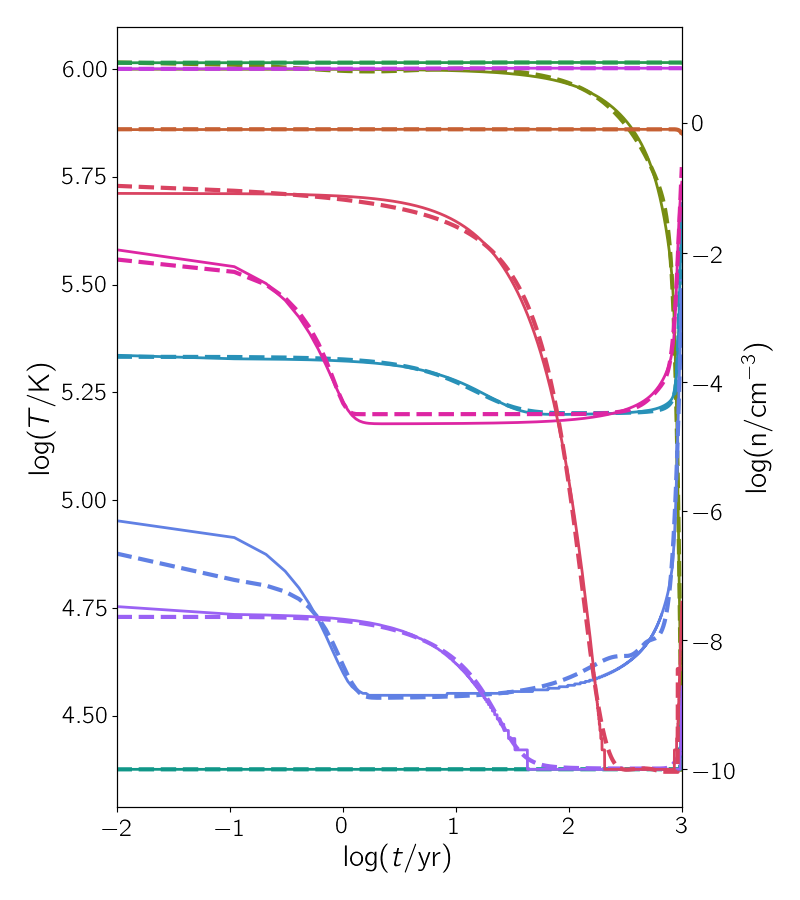}
    \caption{Examples of the time ($t$) evolution of temperature ($T$) and the density ($n$, see the legend) of all the species in the chemical network.
    The solid lines represent the solutions computed using \texttt{KROME}, while the dashed lines depict the predictions of our models, with each line being a single solution from the emulator.
    In the left (right) panel the gas number density is $n=10^4\mathrm{cm}^{-3}$ (initial temperature is $T=10^6\mathrm{K}$), i.e. outside the range of the training dataset (see Tab. \ref{tab:data_set_structure}).
    \label{fig:example_time_evolution_comparison}
    }
\end{figure*}

\begin{figure*}
    \centering
    \includegraphics[width=0.9\textwidth]{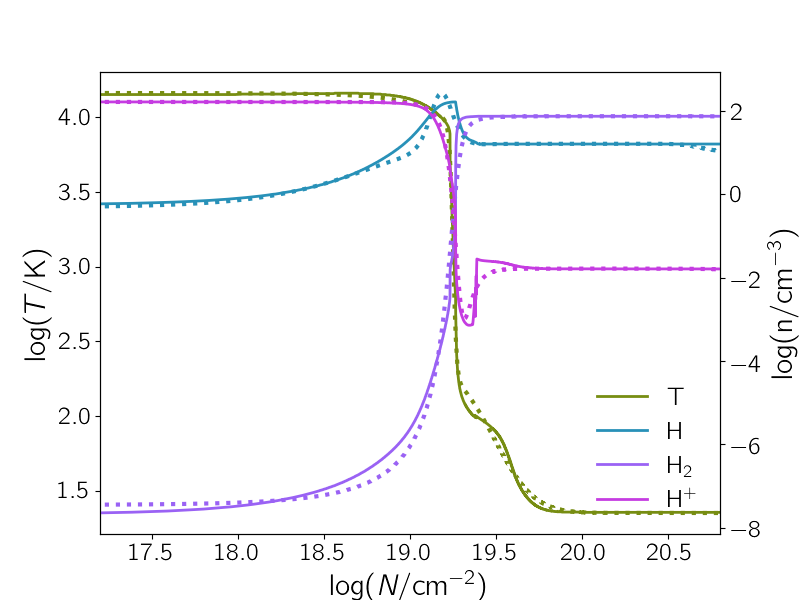}
    \caption{
    Photo-dissociation region (PDR) benchmark.
    Temperature ($T$, right axis) and density of main hydrogen species (H, H$^{+}$, and H$_{2}$, left axis) profiles as a function of column density ($N$) obtained for an impinging radiation flux of $10\,G_0$ propagating in a $Z=\zsun$ slab (see Sec. \ref{sec:pdr_test} for details of the model).
    Solid lines represent the numerical solutions computed with \texttt{KROME} at equilibrium ($t\sim 1\,\mathrm{Kyr}$), dashed lines represent the solution predicted by the emulator.
    Note that each point in the profile at a given $N$ is treated independently by the emulator.
    \label{fig:pdr_multi} 
    }
\end{figure*}

%
% (TRAINING TIME: 15620.524764 s, time prediction: 25.35757040977478 number predictions: 20132656, time krome stimato: 3239.999 s, speed up stimato: 127,772)
%
Here we present the results of our emulator, starting by checking the precision of the model on the training set and validating it outside the training data range (Sec. \ref{sec:testing}), then showcasing the emulator on a physically relevant case of study (Sec. \ref{sec:pdr_test}), and finally comparing with other tools that aim at solving ISM photo-chemistry (Sec. \ref{sec:comparison}).

In this work, each model is trained for a total of $5\times 10^4$ epochs on a single NVIDIA A100 GPU (40GB), for approximately $\simeq 4.34\,\rm GPUhrs$ per each chemical output (temperature and species), i.e. for a total of $\simeq 43.39\,\rm GPUhrs$ for the full set.

\subsection{Model testing and validation}\label{sec:testing}

For the testing, we assessed the robustness of our results by comparing them against a subset of the \texttt{KROME}-generated data, excluded from the training phase.

We start our analysis by showing in Fig. \ref{fig:pred_vs_true} the results from DeepONet vs \texttt{KROME} in the prediction vs true plane, visualized via a bi-dimensional Probability Distribution Function (PDF).
Note that in Fig. \ref{fig:pred_vs_true} we plot the sum of temperature and all the species normalized via eq. \ref{eq:def_sum_ion_and_t}, individual PDFs are reported in Fig.s  \ref{fig:pred_vs_true_temperature} and \ref{fig:pred_vs_true_individual} of App. \ref{sec:app:predictedvstrue}.

For both the normalized version and individual species, the 2D PFDs are strongly peaked around the predicted=true bisector (dashed black line), with a density that decreases by 4 order of magnitude already $\sim 0.1$ dex away from the true solution.
Note that the stripe features seen in the lower left corners of Fig. \ref{fig:pred_vs_true} are due to the sparse sampling of the extrema of the parameter space and is mostly stemming from the summed combination of H, He, and He$^{++}$ (see Fig. \ref{fig:pred_vs_true_individual}).
Qualitatively, this analysis -- in particular the small distance between bisector and PDF peaks in the whole axis -- start to reveal the good accuracy of the predictions as a function of the physical ranges of density and temperature (see Tab. \ref{tab:data_set_structure}).

Quantitative, a summary of the accuracy is provided in Tab. \ref{tab:model_results}, that reports the performance in terms of relative errors defined as
\begin{equation}
\Delta_r = \sqrt{\bigg(\frac{x_{\text{true}}-x_{\text{pred}}}{x_{\text{true}}}\bigg)^2}\,,
\label{eq:error_def}
\end{equation}
where $x_{\text{true}}$ represents values computed with \texttt{KROME}, and $x_{\text{pred}}$ indicates predictions made by our model. Tab. \ref{tab:model_results} shows that the quantitative representation of error distribution for all species has mean relative errors of less than $3\%$, with a median below $2\%$ and a $90\%$ quantile under $6\%$.

To visualize such a benchmark, in Fig. \ref{fig:error_pdf_1D} we plot the PDF of the relative errors for $T$ and all $n_i$.
All the PDFs are peaked around $\Delta_r\sim 10^{-2}$, with the exception for the error distribution of neutral hydrogen, which presents a peak at higher values, but still less than $\Delta_r\sim 10^{-1}$, i.e. 10\%.
In most cases, each PDF resembles a Gaussian, which thus can be characterized by its width; the typical width of the distribution can be expressed in terms of standard deviation, and on average is approximately $\sigma \simeq 1.3$ in $\log(\Delta_r)$ space.
Thus, typically for each species an error of $0.05\% = 10^{-3.3} \lsim \Delta_r\lsim 10^{-0.7}\simeq 0.2$ can be expected for the adopted $\simeq 4.34\,\rm GPUhrs$ of training.

Considering that the training for each species can be performed independently, with the idea of coupling with a simulation it is possible and relatively straightforward to use the $\Delta_r$ PDF (as Fig. \ref{fig:error_pdf_1D}) as a guideline to seek which species can benefit from further training (as likely neutral hydrogen here).
Moreover, it is important to identify the incidence of outliers, which can be defined as cases with a \textit{large} relative error greater than e.g. $\Delta_r>1$. With the current training, such outliers constitute approximately $10^{-6}\%$ of the total, as can be qualitatively inferred from the rapid drop of the PDF in Fig. \ref{fig:pred_vs_true}.
This implies that - for instance - in a numerical simulation with $\sim 10^6$ finite elements, a few outliers are likely to affect the results every $\sim 100$ time step. For the coupling with numerical simulations such outliers should be prevented \citep[see e.g.][]{galligan:2019}. This can be prevented by simply furthering the training (for example with an additional training phase using a second order optimizer, such as L-BFGS) or, more precisely, can be achieved via an anomaly detector \citep{pang:2020}, which should be coupled directly with the DeepONet structure. This will be explored in a future work.

To have a practical idea of the accuracy and performance of the emulator and to validate our model, we illustrate the evolution of chemical species up to $1\mathrm{kyr}$ for a couple of single set of initial conditions.
We can check two different examples.
In the first scenario (left panel of Fig. \ref{fig:example_time_evolution_comparison}) the initial conditions include a total gas density of $n = 10^4 \mathrm{cm}^{-3}$, a temperature of $T = 10^3 \mathrm{K}$, and a (almost negligible) photon flux intensity of $G = 10^{-1} G_0 $.
The second scenario (right panel of Fig. \ref{fig:example_time_evolution_comparison}) features $n = 10^2 \mathrm{cm}^{-3}$, $T = 10^6 \mathrm{K}$, and $G = G_0$.

Note that these examples act as a test since the evaluation has been done at times $t$ which are different with respect to the 16 linearly sampled data points included in the training set.
Moreover, both scenarios can be take as validations, since some species have initial condition outside the training and testing range, i.e. in the first case the gas number density is $n=10^4\mathrm{cm}^{-3}$, in the second $T=10^6\mathrm{K}$ (Tab. \ref{tab:data_set_structure}); 

The results are presented in Fig. \ref{fig:example_time_evolution_comparison}, where $T$ and each $n_i$ are plotted as a function of $t$: with solid lines we report the outcomes computed using the \texttt{KROME} software, with dashed lines we illustrate the predictions from our DeepONet emulator.
The overall evolution is accurately captured by our model as seen by the small, i.e. $\lsim 0.1$ dex, distance between the DeepONet and \texttt{KROME} solutions, with errors that tend to decrease at higher times ($t\simeq 1\, \rm kyr$).
For the second example, some larger discrepancies are present in the early stages (up to approximately $1\, \rm yr$) for He; this inaccuracy reflects the fact that the error distribution of He is among the worst (see Fig. \ref{fig:error_pdf_1D}), i.e. $\Delta_r$ peaks at $10^{-1.7}$. Further, recall that the $t$ training set is extracted from the linear spaced 0 to $1\, \rm kyr$, thus the sampling of the data is expected to be lower at early times. These two facts combined can lead to larger errors in the prediction phase.

Indeed, adopting a linear sampling of the training space in the time domain implies a better sampling at high $t$, thus explains why errors seems to decrease with increasing $t$ for most of the species in both examples.
Moreover, it is to note that our emulator captures the sharp turns of some species very well, for instance for $\mathrm{He}^{++}$ in the left panel and $\mathrm{H}_2^+$ in the right panel, both featuring a numerical gradient of about 10 order of magnitude at $t\sim 10 \rm yr$, hinting at the fact that our method does not suffer too much from the stiffness of the system.

Errors are small despite the initial density and temperature exceeding the maximum values in the training set (Tab. \ref{tab:data_set_structure}), indicating that the model not only interpolates the training data but also seems to gain a good understanding of the influence of initial conditions (the sensors) on the evolution operator.
In general, the problem of extrapolating the solutions of Neural Operators is complex to quantify. \citet{zhu:2023} meticulously analyze the capability of DeepONet to extrapolate solutions, also proposing experimental methods to enhance the adaptability of trained models to receive unexpected inputs, i.e. beyond the initial dataset.
As shown in Figure \ref{fig:example_time_evolution_comparison}, our model seems to generalize quite well within a range about half a dex (both for $n$ and $T$) outside the initial conditions covered by the training set, which can act as a safety insurance if -- for instance during a numerical simulation -- a model reaches an unexpected input value.
This ability to extrapolate outside the range of initial conditions worsens further away from the limits of the training set. For instance, the accuracy decreases by a factor of 10 if we take an initial $T\simeq 10^{7.5} \rm K$, i.e., with respect to the upper bound of the training dataset, outside by a factor of 2 in the normalized log space (see Sec. \ref{sec:deeponet}) .

However, the above discussion does not apply to time extrapolation. In general, it is more challenging for the operator to extrapolate in the time domain. This is primarily due to the non-linear structure of the ODEs system, that can present sharp turns at late time, not included in the testing set for a particular combination of initial conditions.
In principle, it is possible to enhance the time extrapolation capability of DeepOnet by increasing the number of sensors \citep[in particular see Sec. 12.1 in the supplementary material]{lu2021learning}.
Possibly, a more cost-effective approach for longer time integration consists in concatenating the solution of the emulator, using the prediction in the first time step as initial conditions for the second iteration and so on, exploiting the fact that no explicit time dependence is present in the photo-chemical network; however, it is unclear how such a concatenation would affect the error propagation: we leave this analysis for a future work.

\subsection{Photo-dissociation region benchmark}\label{sec:pdr_test}

%
% pdr
%
For a physically relevant benchmark, we simulate a photo-dissociation region (PDR, see \citealt{wolfire:2022} for a review), similarly to the test presented in the photoionization code comparison study from \citet{rollig:2007}.
%
% physical setup
%
Adopting a planar geometry, we take a slab of gas with constant gas density $n= 10^2 \,\cc$ and a maximum column density of $N= 5 \times 10^{21} \rm cm^{-2}$; we assume a constant metallicity $Z = \zsun$ with solar abundances \citep{asplund:2009} and scale linearly the dust content with the solar value.
%
% radiation
%
We set an input radiation field with spectral shape of a black body with temperature $T = 3 \times 10^5 \rm K$, i.e. to mimic a massive star able to efficiently ionize up to He$^{+}$; its intensity is normalized by setting the flux in the Habing band to be $G= 10 \,\gnot$, where $G_{0}=1.6 \times 10^{-3} {\rm erg}\,{\rm cm}^{-2}\,{\rm s}^{-1}$ is the average Milky-Way value \citep{habing:1968}; the radiation field has $10$ frequency bins, one for each of the photo-chemical reaction included in the chemical network (see Sec. \ref{sec:ISM_chemistry}).
%
% numerical setup
%
We split the slab in optically thin cells and allow for the radiation to propagate trough the slab assuming an infinite speed of light; as the radiation propagates, it is absorbed by dust and gas, which chemical composition and temperature is evolved till equilibrium ($t\sim 1 \, \mathrm{kyr}$) by \texttt{KROME}.
Note that the incident spectrum of the photon -- attenuated black body -- can in principle be significantly different with respect to the one used for the training -- constrained extraction of the fluxes -- (see Sec. \ref{sec:dataset}).

In Fig. \ref{fig:pdr_multi} we show the temperature and the main hydrogen ion/molecules profiles as a function of the column density $N$.
It is important to recall that the profiles do not represent a single solution of the system of ODEs, but a collection of different solutions evaluated at the same $t$ by adopting different attenuation for the incident radiation field, which yield a different local radiation field $\mathbf{F}$, and thus different thermodynamic conditions.
The dashed lines representing the predictions of our emulator reproduce well the general trend of the numerical solution from \texttt{KROME} (solid lines), with errors typically of the order of 7\% in the ionized ($N\lsim 10^{18.5}\rm cm^{-2}$) and molecular ($N\gsim 10^{19.5}\rm cm^{-2}$) regions.
However, we see a larger discrepancy in the transition between the ionized and PDR region ($N\sim 10^{19.3}\rm cm^{-2}$), in the order of 30\%, i.e. with errors generally larger than expected from the testing (see Sec. \ref{sec:testing}).

From a numerical point of view, we can interpret such a discrepancy as follows. As noted by \citet[][in particular see the supplementary material, Sec. 18 therein]{lu2021learning} DeepONet by construction respects the Holder condition, which in our case translates into
\begin{equation}\label{eq:holder_condition}
    \left\lVert\mathcal{G}(\rm \mathbf{p}_a)(\mathbf{x})-\mathcal{G}(\rm \mathbf{p}_b)(\mathbf{x})\right\rVert \leq \mathcal{C}\left\lVert \rm \mathbf{p}_a - \rm \mathbf{p}_b \right\rVert\,,
\end{equation}
where $\mathcal{C}$ is a positive scalar and $\rm \mathbf{p}_a$ and $\rm \mathbf{p}_b$ are two different sets of initial conditions and photon fluxes.
The Holder condition (eq. \ref{eq:holder_condition}) implies that the solutions predicted by DeepONet tend to be smooth with respect to changes in the initial data, making it more difficult to predict discontinuities. Stating it differently, if the training set is not sampling finely enough a region of input parameters with large variations of the outputs, the emulator will tend to average out the predictions; a general discussion is more complex, as this feature is due to the phenomenon of neural network smoothness \citep{2019arXiv190511427J}. This discrepancy can likely be ameliorated by having a finer sampled training set in the discontinuity region and/or increasing the architecture size, to improve the expressivity.

From a PDR modeling point of view, note that the discrepancy at the ionized-neutral interface is compatible with the differences between the various photoionization codes adopted in the post-benchmark comparison from \citet[][see in particular the top panel of Fig. 11 therein]{rollig:2007}; such differences are due different implementations and assumptions regarding the photo-chemical rates in the various codes. Thus we can consider as a minor issue the discrepancy we find between DeepONet and the reference \texttt{KROME} solution. Further, adopting a finer grid of initial conditions sampling the ionized-neutral transition and/or expanding the depth of the neural network (currently 6 layers are adopted) should ameliorate the issue.

\subsection{Comparisons with other models}\label{sec:comparison}

% vs PINN
In \citet{branca:2023}, Physics Informed Neural Networks (PINN) are adopted in order to emulate a photo-chemical network very similar to the one in the present; with respect to the results from \citet{branca:2023}, DeepONet achieves a typical accuracy that is about $10 \times$ better with a computational training cost that is about $40 \times$ lower. We consider this to be a stark improvement for the performance; however, training the DeepONet requires pre-computation of a dataset, which in principle can be expensive (in this case, only $\simeq 100\, \rm CPUhr$ are actually used), while PINNs are completely data-independent.
%
% reason for performance increase
The observed performance disparity between the neural operator and the PINN can be attributed to the following factors. The efficacy of the PINN method is known to heavily rely on the initial conditions set selected during training. In \citet{branca:2023}, we developed a technique to generalize for initial conditions that are unknown at the time of training, by elevating the initial condition to a vector. However, there is no strict theoretical guarantee for this approach's feasibility, and we hypothesize that this limitation is the primary cause of the lower accuracy observed in the PINN compared to DeepONet when trained for an equivalent duration.
%
% radiation
In addition to enhanced precision, with respect to the PINN from \citet{branca:2023} a significant advancement is the possibility of having an explicit dependency on incident radiation. This allows the DeepONet model to be effectively coupled with real-time radiative transfer codes, for instance in following the evolution of local molecular clouds \citep{decataldo:2020}. Instead, in order to use DeepONet to study the formation and evolution of galaxies in the EoR \citep{pallottini:2022}, variation of the metallicity should be allowed; we plan to explore this in a future work.

In \citet{grassi:2021} and \citet{sulzer:2023}, the capabilities of autoencoders to compress a large chemical network (29 chemical species and 224 reactions) into a latent space were explored; subsequently, the latent variables were evolved using both standard stiff ODE integrator \citep{grassi:2021} and a neural ODE approach \citep{sulzer:2023}.
The chemical network in these works is larger than the one adopted here (9 chemical species and 52 reactions, Sec. \ref{sec:ISM_chemistry}); this fact naturally favors the compression approach, however in these models the evolution of temperature, the dependence of the evolution from the initial total density, the dependence of the coupling coefficient from temperature, and the impact of incident radiation are not considered; furthermore, another limitation is given by the fact that the size of the latent space is fixed a priori.
Keeping these differences in mind, compared to the present work the overall accuracy is worse by a factor of $\sim 20$ for the case of direct numerical integration in the latent space \citep[][i.e. as for \citealt{branca:2023}]{grassi:2021}, but similar in the case when the neural ODE solver is implemented \citep{sulzer:2023}.

Furthermore, comparing our results on the PDR with those obtained by \texttt{CHEMULATOR} \citep{holdship:2021}, we obtain a better agreement with the numerical solutions, e.g. a $\Delta_r\simeq 7\%$ compared to a $\Delta_r\simeq 20\%$ errors with respect to the reference solutions; however, it is to note our chemical network contains a lower number of species and reactions, i.e. a system with 30 species and 330 reactions is adopted in \citet{holdship:2021}.
Moreover, in addition to smaller relative errors, a notable difference is in the regularity of our solutions, which are much smoother than those from \texttt{CHEMULATOR}. Such a difference is likely driven by the fact that \texttt{CHEMULATOR} is a purely data driven method; instead, DeepONet also tries to reproduce the map between the initial conditions and the family of solutions of the ODE system \citep{lu2021learning}, since it is an application of the UAT for operators \citep{chen2:1995}, which implies DeepONet is expected to follow the Holder condition (eq. \ref{eq:holder_condition}).

% 20,132,656 predictions is 25.36 seconds, which, compared to the 3240.00
Finally, we note that the time required to make $\simeq 2 \times 10^7$ predictions is about $25.36$ CPU seconds; compared to the $3240.0\,\rm CPUs$ needed by \texttt{KROME}, this gives us an approximate speed-up of $\simeq 128$, i.e. effectively making the chemical-evolution task inexpensive.
Such a speed-up with respect to a traditional stiff ODE solver is of the same order of magnitude as the ones reported in \citet{branca:2023} and \citet{grassi:2021}.
However, it is lower by a factor of $\sim 10$ than the one found in \citet{sulzer:2023}, in the case where the integration in the latent space is replaced by a linear fitting function.
In the case of \texttt{CHEMULATOR} \citep{holdship:2021}, the authors report a remarkable speed-up factor of $5 \times 10^4$ compared to \texttt{UCLCHEM} \citep{holdship:2017}, which significantly surpasses our results. However, drawing a direct comparison is challenging as \texttt{UCLCHEM} and \texttt{KROME} are very distinct codes, each employing different underlying ODEs solvers and with a different usage case in numerical simulation, i.e. usually post-processing and on-the-fly usage, respectively.
Indeed, even with the low $128 \times$ speed-up obtained by DeepONet, substituting a traditional ODE solver with our method would effectively make inexpensive the thermo-chemical step in a numerical simulation.

\section{Conclusions}

In this work, we developed a non-equilibrium photo-chemical emulator, in order to cure the computational bottlenecks that hinder the inclusion of detailed chemistry in state-of-the-art astrophysical simulations. This study pioneers the exploration of a Neural Operator, specifically Deep Neural Operator (DeepONet), for this application.

We adopt the InterStellar Medium (ISM) photo-chemical network from \citet[][9 species, 52 reactions]{bovino:2016} that has been used for studies of Giant Molecular Cloud \citep[GMC,][]{decataldo:2019} and galaxies \citep[][]{pallottini:2019}.
We train DeepONet with dataset generated by solving the ISM photo-chemistry via \texttt{KROME} \citep{grassi:2014}, by adopting for the initial conditions a gas number density $n$, temperature $T$, and fractions of each species $n_i/n$ that are sampled in log-spaced range $-2\leq \log(n/\cc) \leq 3.5$, $\log(20) \leq\log(T/\mathrm{K}) \leq 5.5$, and $-6 \leq \log(n_i/n) < 0$, respectively.
Further, we allow for a varying radiation field $\mathbf{F}$ that is composed by 10 energy bins for the photons, with each sampled in the range $F_{i} \in [10^{-15}, 10^{-5}] \mathrm{eV/cm^2/s/Hz}$ by imposing a continuity prior, i.e. such that $|\log(F_{i}/F_{i\pm 1})|\leq 0.15$ in adjacent bins.
Time is sampled in the $0\leq t/\rm kyr \leq 1$ range and the full dataset contains $\simeq 6.4 \times 10^{8}$ models.

The model has been tested and validated both with idealized and physically motivated scenarios, and the key results can be summarized as follows.

\begin{itemize}
    \item[$\bullet$] Our Deep Learning model for ISM chemistry has given results that are both accurate (relative error $\Delta_r\sim 10^{-2}$) and fast to compute ($\sim 128$ times faster than a traditional solver) with a relatively low training time (total of $\simeq 43.4\,\rm GPUhrs$).
    \item[$\bullet$] Employing DeepONet, we have realized significant enhancements over previous models, particularly when compared with Physics Informed Neural Networks (PINNs, \citealt{branca:2023}), with our approach showing a higher accuracy ($10 \times$) at a reduced computational cost ($40 \times$) during the training phase, alongside a streamlined model parameter framework.
    \item[$\bullet$] A critical innovation is the integration of arbitrary radiation fields, a considerable leap beyond the constraints of traditional chemical emulators. This adaptability to diverse radiation fields is a substantial breakthrough, enabling more accurate modeling in scenarios where radiation significantly influences astrophysical dynamics, such as in the dynamics of GMC \citep[e.g.][]{decataldo:2020}.
\end{itemize}
In summary, the present approach seems to surpass the performance of PINNs \citep{branca:2023} in terms of precision and incorporates a direct dependence on radiation fields, outperforming autoencoder-based methods \citep{grassi:2021} for problems with relatively small chemical networks (i.e. 9 species and 52 reactions).

However, a few limitations affect the current emulation.
For instance, we observed larger discrepancies in the transition between ionized and Photo-Dissociation Regions (PDRs), which could be attributed to DeepONet's inherent inclination towards smooth, continuous solutions; while such discrepancy is of the same order of magnitude of the difference between different photoionization codes \citep{rollig:2007}, this characteristic may influence the model's efficacy in predicting abrupt transitions in chemical profiles.
Our model shows robustness in extrapolating beyond the initial condition (density and temperature) ranges, indicating its potential applicability across a broader spectrum of astrophysical scenarios; however, the model's extrapolation capabilities are limited in the temporal domain: this issue can be mitigated by iteratively applying the trained model multiple times, effectively extending its predictive reach. To couple it effectively in a numerical simulation, the propagation of the error should be carefully tested, and outliers (rare, large deviations with respect to the average errors) should be prevented, e.g. via anomaly detection or further training.
An avenue for future enhancement might involve the exploration of multi-output DeepONet models, which could potentially offer increased precision by harnessing inherent conservation laws within the system, thereby addressing the challenges associated with sharp transitions between various astrophysical conditions.

In conclusion, the present work implies a significant leap forward in the modeling of ISM chemistry, offering an emulator with a good balance of precision, versatility, and computational efficiency.
However, addressing the challenges of better managing transitions between distinct regions and refining the model's capability in handling extrapolation beyond the training domain remains a vital area for future research. Such endeavors not only promise to refine the existing model but also pave the way for more comprehensive simulations of complex astrophysical processes.

\begin{acknowledgements}
We gratefully acknowledge computational resources of the Center for High Performance Computing (CHPC) at SNS.
We acknowledge the CINECA award under the ISCRA initiative, for the availability of high performance computing resources and support from the Class C project PINNISM HP10CB99R0 (PI: Branca).
Supported by the Italian Research Center on High Performance Computing Big Data and Quantum Computing (ICSC), project funded by European Union - NextGenerationEU - and the National Recovery and Resilience Plan (NRRP) - Mission 4 Component 2 within the activities of Spoke 3 (Astrophysics and Cosmos Observations)

\end{acknowledgements}

\bibliographystyle{stile/aa_url}
\bibliography{main}     % name of the .bib file

\appendix

\section{Individual predicted vs true tests}\label{sec:app:predictedvstrue}

\begin{figure}
    \centering
    \includegraphics[width=0.49\textwidth]{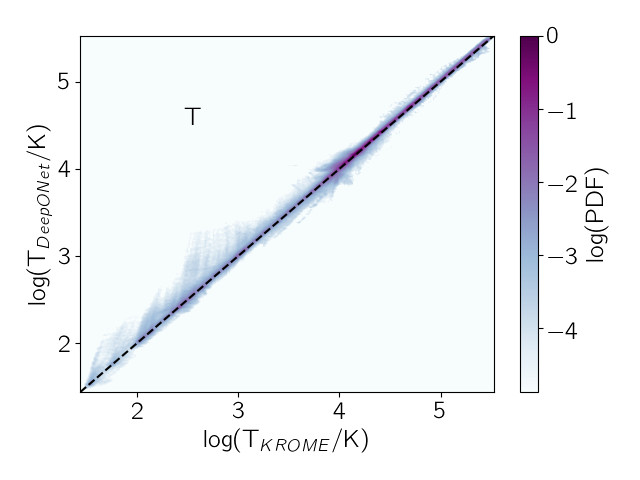} 
    \caption{
    Predicted vs true data shown as 2D PDF for the temperature with respect to \texttt{KROME} data. The black dashed line is the bisector and represent a region without dicrepancy between data and predictions.
    \label{fig:pred_vs_true_temperature}
    }
\end{figure}

\begin{figure*}
    \centering
    \includegraphics[width=0.33\textwidth]{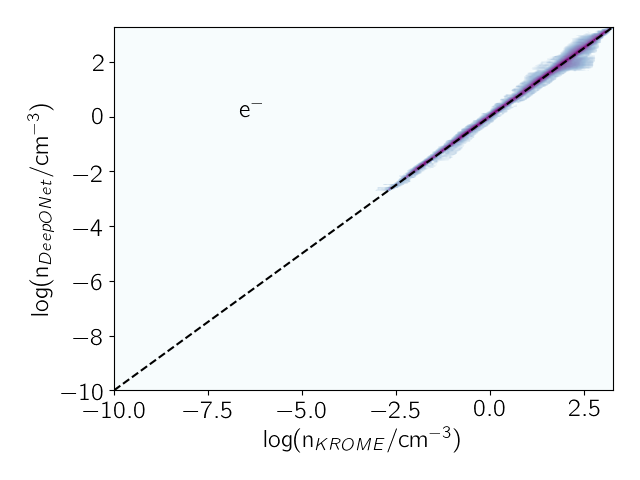}
    \includegraphics[width=0.33\textwidth]{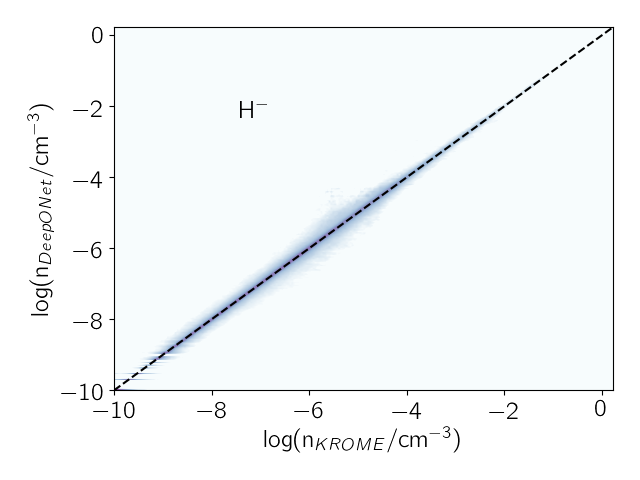}
    \includegraphics[width=0.33\textwidth]{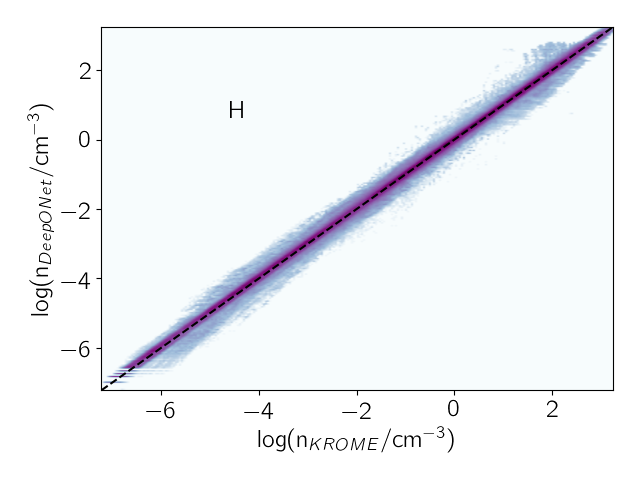}
    
    \includegraphics[width=0.33\textwidth]{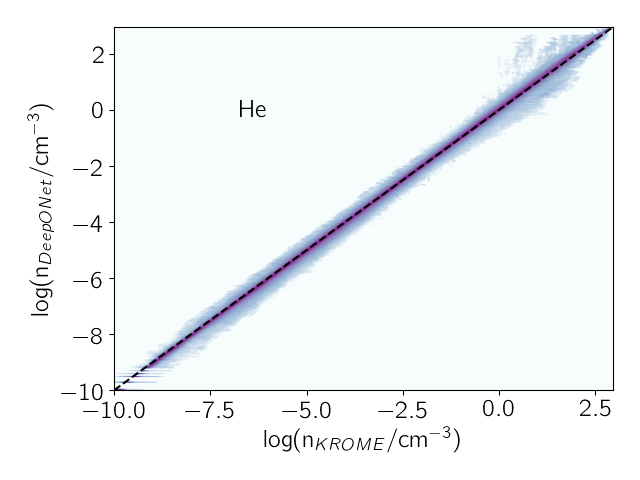}
    \includegraphics[width=0.33\textwidth]{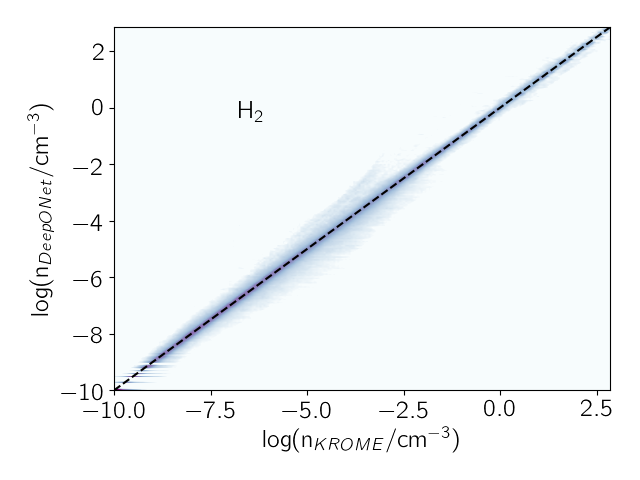}
    \includegraphics[width=0.33\textwidth]{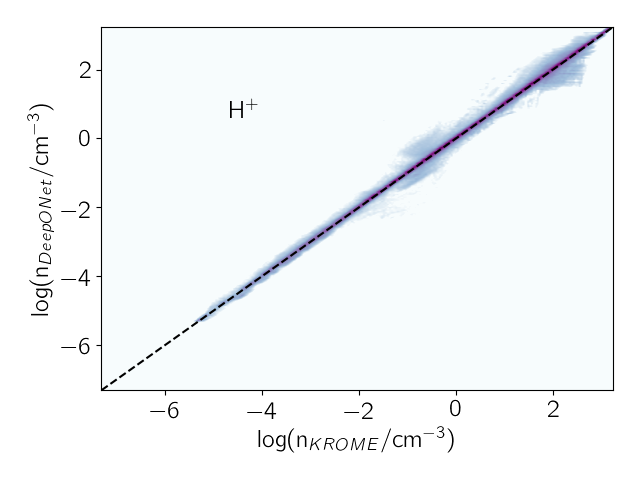}
    
    \includegraphics[width=0.33\textwidth]{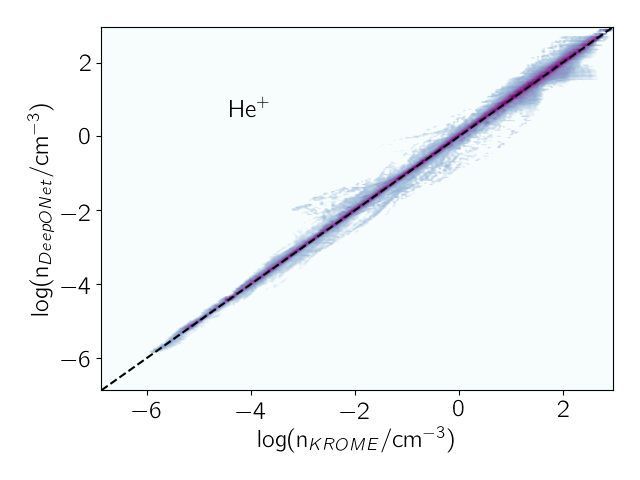}
    \includegraphics[width=0.33\textwidth]{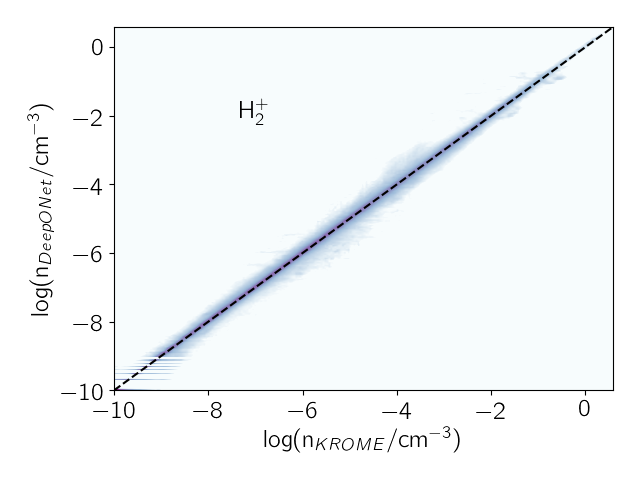}
    \includegraphics[width=0.33\textwidth]{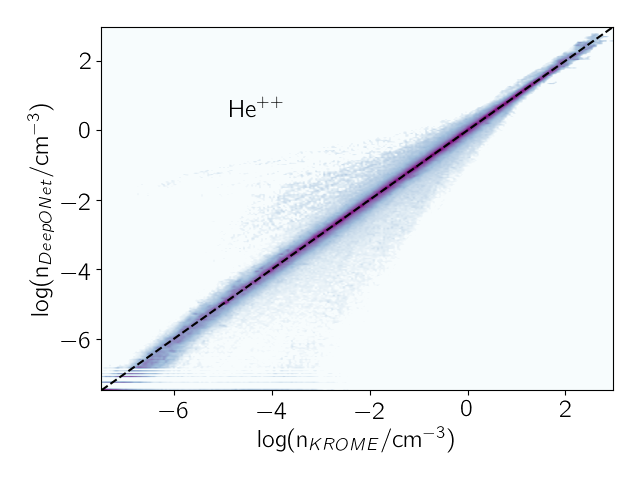}

    \caption{
    Predicted vs true data for all the species with respect to \texttt{KROME} data. The colorbar is given in Fig. \ref{fig:pred_vs_true_temperature}.
    \label{fig:pred_vs_true_individual}
    }
\end{figure*}

In this App., we report the individual predicted vs true analysis for the temperature (Fig. \ref{fig:pred_vs_true_temperature}) and all ions (Fig. \ref{fig:pred_vs_true_individual}) included in the photo-chemical network.
The plot scheme is the same one adopted for the composed analysis shown in Fig. \ref{fig:pred_vs_true} in the main text.

\end{document}